# Angular momentum-dependent topological transport and its experimental realization using a transmission line network

Tianshu Jiang[+], Meng Xiao[+], Wenjie Chen, Lechen Yang, Yawen Fang, Wing Yim Tam and C. T. Chan*

*Department of Physics, Hong Kong University of Science and Technology, Clear Water Bay, Hong Kong, China*

[+]These authors contributed equally to this work.
* Correspondence address: phchan@ust.hk

## Abstract

Novel classical wave phenomenon analogs of the quantum spin Hall effect are mostly based on the construction of pseudo-spins. Here we show that the non-trivial topology of a system can also be realized using orbital angular momentum through angular-momentum-orbital coupling. The idea is illustrated with a tight-binding model and experimentally demonstrated with a transmission line network. We show experimentally that even a very small network cluster exhibits one-way topological edge states, and their properties can be described in terms of local Chern numbers. Our work provides a new mechanism to realize counterparts of the quantum spin Hall effect in classical waves and may offer insights for other systems.

## Introduction

Recent efforts to realize classical wave topological materials have given rise to the field of topological photonics [1-23]. In order to realize the classical counterpart of the quantum Hall effect, the nontrivial band topologies are typically achieved by breaking time reversal symmetry, while the nontrivial topologies of the quantum spin Hall effect (QSHE) are usually realized through spin-orbital coupling. Due to the absence of intrinsic Kramers degeneracy in classical waves, the analogs of the QSHE are realized by constructing pseudo-spins [17-23]. Apart from polarization (spin), the angular momentum of classical waves also offers freedom to control wave [24, 25] and signal propagation [26, 27]. Angular momentum has been treated as a synthetic dimension and the nontrivial topologies made possible by this synthetic dimension have been explored [24,25,28].

Here, we show that a system can exhibit angular momentum-dependent topological properties through angular-momentum-orbital coupling. The boundary of such a system possesses one-way edge states that are locked to the angular momentum without breaking time reversal symmetry. We also provide a proof-of-principle experimental demonstration using a transmission line network. We will see that local Chern numbers [29, 30] can be used to characterize the topology of a small cluster of such network systems. For simplicity, we limit our discussion to a hexagonal network in this work, but the ideas can be easily generalized to other systems.

## Results

In 2D systems, electromagnetic waves decouple into two independent transverse electric and transverse magnetic modes whose evolution can then be represented by a scalar component as denoted by $\varphi$. Consider a cylindrical meta-atom whose eigenfield ($H_z$ or $E_z$) can be written as

$$\varphi(r,\theta) = \varphi_0(r)\exp(im\theta), \tag{1}$$

where $r$, $\theta$ are the polar coordinates, and $m$ denotes the orbital angular momentum. The phase distribution of such a meta-atom with angular momentum $m=1$ is shown in Fig. 1a. If the angular momentum eigenmodes in the system do not interact (which is usually ensured by the rotational symmetry), the system Hamiltonian can be block-diagonalized and each block is labeled by the corresponding angular momentum eigenvalue. Meta-atoms that possess a well-defined angular momentum can be regarded as elementary building blocks of the system and they can form a lattice structure. One such lattice structure is shown pictorially in Fig. 1b, where a 2D honeycomb lattice is considered. Below we will show that periodic systems and finite-size clusters comprising such meta-atoms exhibit topological properties by introducing angular-momentum-orbital coupling.

**Tight-binding model.** To put our idea in context, we first consider a simple tight-binding model which is periodic in the *xy* plane. A meta-atom which exhibits well-defined angular momentum eigenmodes can be realized with a discrete set of *N* nodes (illustrated in Fig. 1c) uniformly spaced in a ring, where *N* is the total number of nodes. These nodes have the same wave amplitude and the phase of the *j*-th node is $j2\pi m/N$. These nodes need not lie on the *xy* plane and in fact, they will be stacked in the *z* direction in the following discussion. Hence Fig. 1c should be regarded as their projected positions on a plane. Such a ring of nodes can obviously exhibit *N* different values of angular momentum. When $N=1$ or $2$, the subspace of each angular momentum still possesses time reversal symmetry. For a system with time reversal symmetry to exhibit nontrivial topology, *N* must be larger than 2. A discretized example for *m*=1 and *N*=3 is shown in Fig. 1d. Similarly, a hexagonal lattice of meta-atoms with angular momentum *m*=1 in Fig. 1b can be represented by discretized nodes as shown in Fig. 1e. For illustration purpose, we label each node with a layer number and the honeycomb lattice in Fig. 1e means that the nodes in each layer form a honeycomb lattice with the same hopping strength as shown in Fig. 1f with the bonds in blue. The gray spheres in Fig. 1f represent the nodes which are assumed to be identical and hence the on-site energies are all set to zero. Each honeycomb lattice consists of two sublattices and hence there are 2*N* nodes in each unit cell.

We now proceed to introduce the angular-momentum-orbital coupling. Such couplings essentially help distinguish different angular momentums and hence modes with different angular momentum eigenvalues experience different synthetic gauge fields. One such coupling is shown by the bonds in yellow in Fig. 1f. Such a coupling introduces a chiral coupling to the AA stacked honeycomb lattice. We note that chiral coupling has been used in constructing Weyl semimetals [31,32]. The angular momentum is preserved as long as the couplings between layers *j* and *j*+1 (layers *N* and 1 when *j*=*N*) are the same for arbitrary values of *j*. Interlayer coupling is introduced between the first layer and the *N*-th layer to preserve the angular momentum, i.e., the *N*-th layer is connected from head to tail to the first layer. The schematic of the tight-binding model for *N*=3 is presented in Fig. 1g. In this *N*=3 example, the nodes $A_1$, $A_2$ and $A_3$ stacked along the z-axis constitute one meta-atom on the sublattice *A* and the nodes $B_1$, $B_2$ and $B_3$ form another meta-atom on sublattice *B*.

As discussed before, the system Hamiltonian can be block-diagonalized with each block having a different angular momentum. We define the angular momentum basis as

$$|m\rangle = \frac{1}{\sqrt{N}} \sum_{j=1}^{N} e^{i2m\pi(j-1)/N} |j\rangle, \qquad (2)$$

where $|j\rangle$ in the summation denotes the original basis for the nodes in the *j*-th layer. In the angular momentum basis, the Hamiltonian can be block-diagonalized and each angular momentum block can be written as (see Supplementary Note 1 for more details)

$$H(\mathbf{k}) = f_x(\mathbf{k})\sigma_x + f_y(\mathbf{k})\sigma_y + f_0(\mathbf{k})\cos(2m\pi/3)\sigma_0 + f_z(\mathbf{k})\sin(2m\pi/3)\sigma_z, \qquad (3)$$

where $\mathbf{k} = (k_x, k_y)$ is the Bloch wavevector, $\sigma_0$ is the $2\times 2$ identity matrix, $\sigma_x$, $\sigma_y$, and $\sigma_z$ are the Pauli matrices, and

$$\begin{aligned}
f_0(\mathbf{k}) &= t_1[4\cos(\tfrac{3}{2}k_x a)\cos(\tfrac{\sqrt{3}}{2}k_y a) + 2\cos(\sqrt{3}k_y a)], \\
f_x(\mathbf{k}) &= t_2[1 + 2\cos(\tfrac{\sqrt{3}}{2}k_y a)\cos(\tfrac{3}{2}k_x a)], \\
f_y(\mathbf{k}) &= 2t_2\cos(\tfrac{\sqrt{3}}{2}k_y a)\sin(\tfrac{3}{2}k_x a), \\
f_z(\mathbf{k}) &= t_1[4\cos(\tfrac{3}{2}k_x a)\sin(\tfrac{\sqrt{3}}{2}k_y a) - 2\sin(\sqrt{3}k_y a)],
\end{aligned} \qquad (4)$$

where $a$ is the lattice constant. Here $t_1$ and $t_2$ denote the interlayer and intralayer hopping strength as represented by the bonds in yellow and blue in Fig. 1f, respectively. In a system with time reversal symmetry, such as the one we are considering, $t_1$ and $t_2$ are both real numbers. The first two terms in Eq. (3) are the same as those in the tight-binding model of graphene [33], which is not surprising as we have a honeycomb lattice. The last two terms in Eq. (3) represent the angular-momentum-orbital coupling, which depends explicitly on the angular momentum. The third term introduces a global energy shift only, and so it does not change the topology of the band structure. The last term introduces an angular momentum-dependent mass term which lifts the Dirac cone degeneracy at K and K' for nonzero *m*. To see this more clearly, we expand the Hamiltonian around K $\left(0, 4\pi/3\sqrt{3}a\right)$ and K' $\left(0, -4\pi/3\sqrt{3}a\right)$ as follows:

$$H(\mathbf{p}) = \frac{9}{2}t_1\left(|m| - \frac{2}{3}\right)\sigma_0 - \frac{3}{2}at_2\left(p_x\sigma_y + \tau p_y\sigma_x\right) + \frac{9}{2}t_1 m\tau\sigma_z, \qquad (5)$$

where $\tau = 1$ for the K point and $\tau = -1$ for the K' point, $\mathbf{p} = (p_x, p_y)$ is momentum measured from the *K* or *K'* point. The last term in Eq. (5) represents the angular-momentum-orbital coupling which is similar to the spin-orbital coupling [34,35]. There is, however, a sector with $m = 0$ where the mass term vanishes, wherein the bands are degenerate at the K (K') point as guaranteed by time reversal symmetry. Although the whole system is invariant under time reversal, the band topology of the $m = \pm 1$ sectors can still be nontrivial as each sector on its own is not time reversal invariant.

**Transmission line network.** We proceed to implement the above tight-binding model with a transmission line network [36-39], which provides the flexibility to realize the "head to tail" connection. The transmission lines here are coaxial cables which can be regarded as one-dimensional waveguides. These coaxial cables are then connected at nodes to form a network, which is described by a set of equations as follows [36]:

$$-\psi_i \sum_j \coth(gl_{ij}) + \sum_j \frac{1}{\sinh(gl_{ij})} \psi_j = 0 \ . \tag{6}$$

Here, $\psi_i$ is the voltage at the $i$-th node, $l_{ij}$ is the length of the cable connecting nodes i and j, and $g = (i\omega/c_0)\sqrt{\varepsilon}$ with $\omega$, $c_0$ and $\varepsilon$ being the angular frequency, the speed of light in vacuum, and the relative permittivity of the dielectric medium in the coaxial cables, respectively. This network equation is equivalent to a tight-binding model with an on-site term $-\sum_j \coth(gl_{ij})$ and a hopping term $1/\sinh(gl_{ij})$. The connecting nodes can be regarded as the discrete nodes considered previously and a cable provides coupling between two nodes. The coupling coefficient depends on the frequency, length and permittivity of the cable. The network topology provides the flexibility to connect the first layer back to the third layer and hence realize the Hamiltonian in Eq. (3). We note that although the hopping here depends on the frequency, this frequency dependence does not change the topology of the system. The idea discussed here can also be extended to other waveguide networks.

**Bulk band and one-way edge states.** To show that the transmission line network exhibits the basic characters of the tight-binding model, we solve Eq. (6) numerically and plot the band structures in Figs. 2a and 2b with $m = 0$ and $m = 1$, respectively. The coaxial cables are connected in the way shown in Fig. 1g and periodic boundary conditions are applied in the in-plane directions. The lengths of intralayer and interlayer cables are taken to be $a$=0.43 m and $b$=2.06 m, respectively. The wave speed inside the cable is assumed to be 0.66c (the same as experimentally measured results). For simplicity, the loss in the cable is ignored for now. The band structures in Figs. 2a and 2b are quite similar to those of the tight-binding model as shown in Supplementary Figs. 1a and 1b and

**Supplementary Note 1.** The band structure of $m=-1$ is the same as that of $m=1$. For $m=0$, the Dirac point appears at the K point as predicted by the tight-binding Hamiltonian where there is no $\sigma_z$ term. For $m=1$, a band gap emerges between two bulk bands. Our system is $C_6$ invariant and the Chern number of each band can be obtained using the rotational eigenvalues at high symmetry points [40,41]. The Chern number of each m=1 band is labeled in Fig. 2b (the Chern numbers have opposite signs when $m=-1$). We can see that a nontrivial band gap exists from 31 MHz to 36.3 MHz. We also show the projected band structure along the *x* direction of the *m*=1 sector with the armchair boundaries in Fig. 2c, where the gray area represents the projection of bulk bands, and the red and blue curves represent the edge states localized at the upper and lower boundaries of a strip, respectively. The existence of these one-way edge states verifies once again the nontrivial topology of the band structure for the *m*=1 sector.

**Local Chern number.** The Chern number in Fig. 2 is obtained by considering an infinite system. Next, we explore the topological property of a finite-size cluster. We consider a sample which is of finite size in the xy plane while the "head to tail" connection is kept such that the angular momentum remains well defined. To study the topological characteristics of a finite-size structure, we adopt the concept of the local Chern number [29,30], which is defined by the anti-symmetric product of the projection operators:

$$\nu(P) = 12\pi i \sum_{j \in A} \sum_{k \in B} \sum_{l \in C} (P_{jk} P_{kl} P_{lj} - P_{jl} P_{lk} P_{kj}), \tag{7}$$

where $P = \sum_{f \leq f_c} |u_f\rangle\langle u_f|$ is the projection operator which adds up all the eigenstates $|u_f\rangle$ below a cutoff frequency $f_c$, and $P_{ij} = \langle x_i | P | x_j \rangle$ is the spatial relation between sites $x_i$ and $x_j$. These sites lie in three different sectors (labeled *A*, *B*, and *C* in the counterclockwise direction) of the circular computational domain (see left panel of Fig. 3a). The cutoff frequency $f_c$ is set to 31 MHz, which is the lower band edge frequency of the nontrivial band gap. Figure 3a shows a top view of the finite-size 9x9 lattice with the computational domain at the center. Each sector (A, B, or C) covers $120°$ and the computational domain is a circle with radius *r*. For the sites on the boundaries between two sectors, we distribute them to the adjacent sector in the clockwise direction. The value of $\nu$ as defined in Eq. (7) for such a chosen computational domain is taken as the local Chern number at the center of this circular region. The radius *r* of this computational region will affect the value of $\nu$. To explore this dependence, we plot the $\nu$-$r$ curve in the right panel of Fig. 3a for the $9 \times 9$ sample

shown in the left panel, where the center of the circle is fixed at the center of the sample. From the results shown in Fig. 2b, we expect the local Chern number to be $-1$ (1) for $m=1$ ($m=-1$). We can see that $\nu$ goes to zero when $r$ is too small or too large compared with the sample size, but when the radius takes intermediate values, the result converges to -1 (1) for $m=1$ ($m=-1$), consistent with the Chern numbers calculated under periodic boundary conditions. In the following calculation, the radii of the computational region are set to $2.3a$ and $5a$ for the $3\times 3$ and $9\times 9$ samples, respectively.

By solving Eq. (6), we obtain the eigen-spectrum for $m=1$ as shown in the left panel of Fig. 3b, where each black dashed line corresponds to the frequency of an eigenstate. The local Chern number $\nu$ as a function of the cut-off frequency (the summation is taken over all of the eigenstates below this frequency) is shown in the middle panel of Fig. 3b when the center of the sample is taken as the center of the computational region. We see that there is a wide spectral range where $\nu$ is almost $-1$. This spectral range starts from about 31 MHz and ends near 36 MHz, which coincides with the lower and upper edges of the nontrivial band gap as determined by the band structure calculation. We next set the cutoff frequency to 31 MHz (marked by the red dotted line in the left panel of Fig. 3b), and calculate $\nu$ as a function of the center position of the computational region. The corresponding results are shown in the right panel of Fig. 3b. We see that except for the boundary region where the blue color is a lighter shade, other parts exhibit a local Chern number close to -1. Such a nontrivial local topological property persists even when the sample size decreases to $3\times 3$ as shown in Fig. 3c. Once again, the left, middle and right panels show the eigen-spectrum, local Chern number at the center, and local Chern number as a function of position, respectively. Here the cutoff frequency is also chosen as 31 MHz.

**Experimental demonstration of the one-way edge states.** Due to the nontrivial topological property of the $m=\pm 1$ subspace and the existence of the local Chern number, finite-size samples should have one-way edge states. The robustness of these one-way edge states against defects and sharp corners are numerically investigated in Supplementary Fig. 2 and Supplementary Note 2. Due to the complexity of the connected network, we work with a small sample. As shown in the previous section, the nontrivial topology should be manifested in the transmission spectra even for a $3\times 3$ sample. The experimental sample is shown in Fig. 4a. Fig. 4b shows the connectors in our system. One node (left hand side of the panel) consists of four cross connectors (upper-right) and three straight connectors (lower-right), connecting intralayer, interlayer and measurement cables. In our experiments, we use two cables of the same type (RG58C/U type [42]) but with different lengths as shown in Fig. 4c. The one for interlayer connections is around 2.00 m long; the one for intralayer connections is around 0.37 m long. A hexagon formed by the connectors and cables is shown in Fig. 4d. Fig. 5a shows the in-plane lattice structure of the experimental sample in Fig. 4a and the interlayer coupling is the same as that shown in Fig. 1g. The parameters of the lattice are the same as those in Fig. 2. We also add additional cables in the upper-right corner (indicated by the red line) as a defect to test the one-way property of the edge states. (More experimental details can be found in Methods.) The transmission spectra of the system with and without the defect are shown in Fig. 5b with the red and black curves, where the gray region denotes the nontrivial gap region. It can be seen that inside the nontrivial gap region, the transmission spectra are almost the same, while they differ for frequencies below the gap region. As the scattering of the defect is weak above the gap region, the difference between these two transmission spectra is small.

We also measure the field distribution of the edge state excitation as additional evidence of the existence and robustness of the one-way edge states. Figs. 5c and 5d show respectively the experimental measurement and numerical simulation of the field distribution at 34.5 MHz for $m=1$. As the cables in the experiments have intrinsic loss, we also add intrinsic loss obtained from fitting the experimental data to the numerical simulation (see Methods and Supplementary Fig. 3). The loss in the cable used in the experiment is rather small, and $g \approx ik - 1/2L$, where $k = \omega\sqrt{\varepsilon'}/c_0$ and absorption length $L = \varepsilon'/k\varepsilon''$ with $\varepsilon'$ and $\varepsilon''$ being the real and imaginary parts of the relative permittivity. In this study, the absorption length of our cables is measured to be $L \cong 338 \cdot f^{-0.6123}$, where $L$ is in meters (m), and $f$ is in MHz. The experimental results agree well with the numerical results. It is clear that the fields are localized in the boundary layer and propagate unidirectionally, uninterrupted by the defect. The voltage magnitude is attenuated along the transportation direction because of the loss in the cable. The propagation direction is locked to the orbital angular momentum, and for the $m=1$ sector shown in Fig. 5, the one-way edge states propagate in a clockwise manner. For the $m=-1$ sector, the edge states propagate in the anti-clockwise manner (Supplementary Figs. 6a-6d). We also measure the field patterns for $m=0$. The wave propagates at 33.7 MHz inside the bulk band (Supplementary Fig. 5a) but is reflected at 49.2 MHz which falls inside the band gap (Supplementary Fig. 5b). For $m=0$, we do not observe any nontrivial edge states at any frequency as it is topologically trivial.

## Discussion

We have demonstrated that angular momentum can provide an additional degree of freedom to control the topology of a system through coupling with orbital momentum. Such an idea is illustrated with a tight-binding model and experimentally verified with a transmission line network. The well-defined local Chern number shows that the topological properties persist even for a very small cluster, as was realized experimentally. We experimentally realized the $N=3$ case but samples with higher values of $N$ can be constructed and in principle each angular momentum subspace can carry a different topology. Our idea can be generalized to other waveguide network systems. Moreover, such an idea is not limited to the discretized model, and for a network system, it can be extended to higher dimensions.

## Methods

**Experimental materials**

The cables we use have intrinsic loss, whose values are calibrated experimentally. Due to the intrinsic loss, the wave amplitude along the propagation direction will attenuate at the rate of $\exp(-x/2L)$, where $x$ is the propagation distance and $L$ is the absorption length. The absorption length is frequency dependent and can be described by an empirical relation $L \cong \alpha f^{-\beta}$ in the frequency range of our experiments [36]. Here $f$ represents frequency in MHz, and $\alpha$ and $\beta$ are constants determined from fitting the experimental data. In Supplementary Fig. 3a, we show the log of the field amplitudes at different frequencies as functions of cable lengths, and their linear fitting gives the absorption length at each frequency. The log of the absorption lengths are then shown in Supplementary Fig. 3b (as dots) and are fitted to obtain $\alpha = 338\,\text{m}$ and $\beta = 0.6123$. The wave speed of this cable is 0.66c, and its frequency dependence can be safely ignored, especially as the loss is quite low. As evidence, the imaginary part of the relative permittivity at 30 MHz is estimated to be about 0.06, which is much smaller than the real part at approximately 2.3.

**Experimental details**

In the experiment, we measure the property inside each angular momentum sector only, which can be ensured if the source only excites the angular momentum of that sector. In Supplementary Fig. 4a, we show one such setup for $N=3$, where $m$ in the equation determines the angular momentum to excite. Taking advantage of the superposition principle, we excite each layer separately and then add up the amplitudes (the phase delay due to the angular momentum is included) to obtain the field distribution. In the case where only one angular momentum sector is excited, the field amplitudes of the layers differ only by a phase factor. Hence in all the plots of field distributions, we only show the field amplitude of one layer.

Due to the presence of the connectors, the length of the cables used is slightly shorter than the actual distance between two nodes. To obtain the actual distance between two nodes, we use the connection as shown in Supplementary Fig. 4b. Same as before, we only show the connection within one layer. The interlayer connections are the same as before and periodic boundary conditions are applied to the left and right boundaries. The source is at the node at the upper edge and we measure the transmission at the node at the lower edge as shown in Supplementary Fig. 4b. In such a setup, we actually excite the band along the $k_x = 0$ direction as highlighted in red in Supplementary Fig. 4c. The extra length added due to the presence of the connectors is the same for all cables. The blue curve in the left panel of Supplementary Fig. 4d shows the experimentally measured transmission spectrum, while the red curve represents the numerically calculated transmission spectrum with the extra length being 0.06 m. These two transmission spectra match quite well. The right panel of Supplementary Fig. 4d shows the corresponding band structure of *m*=1 along the $k_x = 0$ direction. Due to the limited size of the sample used in the experiment, the transmission inside the band gap (gray region) is shallow but can still be seen clearly. In Supplementary Fig. 4d, we also show the numerically calculated transmission spectrum with 20 unit cells along the *y* direction with the cyan curve. Here, we increase the magnitude of the signal by five times. With more unit cells, the band gap frequencies continue to show no transmittance as expected. The transmission amplitude also decreases and the spectrum becomes smoother due to the intrinsic loss in the cable.

## References

1. Haldane, F. & Raghu, S. Possible realization of directional optical waveguides in photonic crystals with broken time-reversal symmetry. *Phys. Rev. Lett.* **100,** 013904 (2008).
2. Raghu, S. & Haldane, F. D. M. Analogs of quantum-Hall-effect edge states in photonic crystals. *Phys. Rev. A* **78,** 033834 (2008).
3. Wang, Z., Chong, Y., Joannopoulos, J. & Soljačić, M. Reflection-free one-way edge modes in a gyromagnetic photonic crystal. *Phys. Rev. Lett.* **100,** 13905 (2008).
4. Wang, Z., Chong, Y., Joannopoulos, J. D. & Soljačić, M. Observation of unidirectional backscattering-immune topological electromagnetic states. *Nature* **461,** 772–775 (2009).

5. Lu, L., Joannopoulos, J. D. & Soljačić, M. Topological photonics. *Nat. Photon.* **8,** 821–829 (2014).

6. Xiao, M. & Fan, S. Photonic Chern insulator through homogenization of an array of particles. *Phys. Rev. B* **96,** 100202(R) (2017).

7. Ozawa, T., Price, H. M., Amo, A., Goldman, N., Hafezi, M., Lu, L., Rechtsman, M., Schuster, D., Simon, J., Zilberberg, O., & Carusotto, I. Topological photonics. Preprint found at https://arxiv.org/abs/1802.04173 (2018).

8. Kraus, Y. E., Lahini, Y., Ringel, Z., Verbin, M. & Zilberberg, O. Topological states and adiabatic pumping in quasicrystals. *Phys. Rev. Lett.* **109,** 106402 (2012).

9. Rechtsman, M. C. et al. Photonic Floquet topological insulators. *Nature* **496,** 196–200 (2013).

10. Lu, L., Joannopoulos, J., D. & Soljačić, M. Topological states in photonic systems. *Nat. Phys.* **12,** 626-629 (2016).

11. Yuan, L., Shi, Y. & Fan, S. Photonic gauge potential in a system with a synthetic frequency dimension. *Opt. Lett.* **41**, 741-744 (2016).

12. Fang, K., Yu, Z. & Fan, S. Realizing effective magnetic field for photons by controlling the phase of dynamic modulation. *Nat. Photonics* **6**, 782-787 (2012).

13. Yu, Z., Veronis, G., Wang, Z. & Fan, S. One-way electromagnetic waveguide formed at the interface between a plasmonic metal under a static magnetic field and a photonic crystal. *Phys. Rev. Lett.* **100**, 023902 (2008).

14. Ao, X., Lin, Z. & Chan, C. T. One-way edge mode in a magneto-optical honeycomb photonic crystal. *Phys. Rev. B* **80,** 033105 (2009).

15. Poo, Y., Wu, R., Lin, Z., Yang, Y. & Chan, C. T. Experimental realization of self-guiding unidirectional electromagnetic edge states. *Phys. Rev. Lett.* **106,** 093903 (2011).

16. Chen, X. D., Deng, Z. L., Chen, W. J., Wang, J. R. & Dong, J. W. Manipulating pseudospin-polarized state of light in dispersion-immune photonic topological metacrystals. *Phys. Rev. B* **92,** 014210 (2015).

17. Hafezi, M., Demler, E. A., Lukin, M. D. & Taylor, J. M. Robust optical delay lines with topological protection. *Nat. Phys.* **7,** 907–912 (2011).

18. Hafezi, M., Mittal, S., Fan, J., Migdall, A. & Taylor, J. M. Imaging topological edge states in silicon photonics. *Nat. Photon.* **7,** 1001–1005 (2013).

19. Khanikaev, A. B. et al. Photonic topological insulators. *Nat. Mater.* **12,** 233–239 (2013).

20. Cheng, X. et al. Robust reconfigurable electromagnetic pathways within a photonic topological insulator. *Nat. Mater.* **15,** 542–548 (2016).

21. Chen, W.-J. et al. Experimental realization of photonic topological insulator in a uniaxial metacrystal waveguide. *Nat. Commun.* **5,** 5782 (2014).

22. Yang, Y. et al. Visualization of a unidirectional electromagnetic waveguide using topological photonic crystals made of dielectric materials. *Phys. Rev. Lett.* **120**, 217401 (2018).

23. Wu, L.-H. & Hu, X. Scheme for achieving a topological photonic crystal by using dielectric material. *Phys. Rev. Lett.* **114,** 223901 (2015).

24. Luo, X.-W. et al. Quantum simulation of 2D topological physics in a 1D array of optical cavities. *Nat. Commun.* **6,** 7704 (2014).

25. Luo, X.-W. et al. Synthetic-lattice enabled all-optical devices based on orbital angular momentum of light. *Nat. Commun.* **8**, 16097 (2017).

26. Wang, J. et al. Terabit free-space data transmission employing orbital angular momentum multiplexing. *Nat. Photon.* **6**, 488 (2012).

27. Bozinovic, N. et al. Terabit-scale orbital angular momentum mode division multiplexing in fibers. *Science* **340**, 1545 (2013).

28. Ozawa, T., Price, H. M., Goldman, N., Zilberberg, O. & Carusotto, I. Synthetic dimensions in integrated photonics: from optical isolation to 4D quantum Hall physics. *Phys. Rev. A* **93,** 043827 (2015).

29. Mitchell, N. P., Nash, L. M., Hexner, D., Turner, A. & Irvine, W. T. M. Amorphous topological insulators constructed from random point sets. *Nat. Phys.* **14,** 380–385 (2018).

30. Kitaev, A. Anyons in an exactly solved model and beyond. *Ann. Phys. (Leipz.)* **321,** 2–111 (2006).

31. Xiao, M., Chen, W., He, W. & Chan, C. T. Synthetic gauge flux and Weyl points in acoustic systems. *Nat. Phys.* **11,** 920–924 (2015).

32. Chen, W.-J., Xiao, M. & Chan, C. T. Photonic crystals possessing multiple Weyl points and the experimental observation of robust surface states. *Nat. Commun.* **7,** 13038 (2016).

33. Peres, N. M. R. The transport properties of graphene: an introduction. *Rev. Mod. Phys.* **82,** 2673–2700 (2010).

34. Kane, C. L. & Mele, E. J. Quantum spin hall effect in graphene. *Phys. Rev. Lett.* **95,** 226801 (2005).

35. Kane, C. L. & Mele, E. J. Z2 topological order and the quantum spin Hall effect. *Phys. Rev. Lett.* **95,** 146802 (2005).
36. Zhang, Z. Q. et al. Observation of localized electromagnetic waves in three-dimensional networks of waveguides. *Phys. Rev. Lett.* **81**, 5540 (1998).
37. Li, M., Liu, Y. & Zhang, Z. Q. Photonic band structure of Sierpinski waveguide networks. *Phys. Rev. B* **61**, 16193 (2000).
38. Zhang, Z. Q. & Sheng, P. Wave localization in random networks. *Phys. Rev. B* **49**, 83 (1994).
39. Cheung, S. K., Chan, T. L., Zhang, Z. Q. & Chan, C. T. Large photonic band gaps in certain periodic and quasiperiodic networks in two and three dimensions. *Phys. Rev. B* **70**, 125104 (2004).
40. Fang, C., Gilbert, M. J. & Bernevig, B. A. Bulk topological invariants in noninteracting point group symmetric insulators. *Phys. Rev. B* **86,** 115112 (2012).
41. Fang, C., Gilbert, M. J., Dai, X. & Bernevig, B. A. Multi-Weyl topological semimetals stabilized by point group symmetry. *Phys. Rev. Lett.* **108,** 266802 (2012).
42. https://hkcn.rs-online.com/web/p/thin-ethernet-cable/5218436/?sra=pstk

## Acknowledgements

This work was supported by the Research Grants Council of Hong Kong (16304717 and AoE/P-02/12).

## Author contributions

C. T. C. initiated the program and oversaw and directed the entire project. M.X. conceived the idea of controlling waves with angular momentum. T.-S.J. developed the theory and carried out the experiment. W.-J.C. provided important suggestions for theory and experiment. L.-C.Y., Y.-W. F. and W. Y. T. provided experimental support. All authors contributed to the analysis and the discussion of results.

## Competing financial interests

The authors declare no competing financial interests.

# Figures

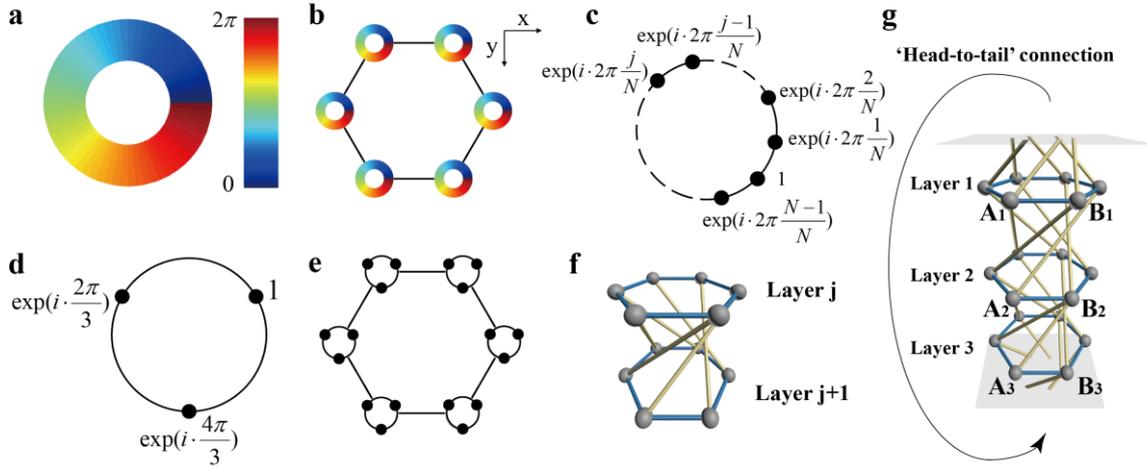

**Figure 1| Realization of angular-momentum-orbital coupling in 2D honeycomb lattices.** (**a**) Phase distribution of a mode with angular momentum m=1. (**b**) A honeycomb lattice with sites exhibiting nonzero angular momentum. (**c**) A discrete set of N point nodes, where each carries a different phase, can emulate a mode with nonzero angular momentum (**d**) A minimum of three nodes is needed to create a nontrivial topology. (**e**) A discretized version of the honeycomb lattice in (**b**), with each site carrying three nodes. The three nodes need not lie geometrically on the same plane. (**f**) An exemplary connection which exhibits nontrivial angular-momentum-orbital coupling. Here a layer represents the lattice structure formed by the nodes with the same sequence number *j* (as shown in (**c**)) of different loops on each lattice site. Gray spheres represent nodes. The bonds in blue and yellow indicate intralayer and interlayer couplings, respectively. (**g**) An *N*=3 example. The 'head-to-tail' connection here means layer 1 is connected to layer 3 via the same connections as those shown in (**f**). The loop formed by the nodes $A_1$, $A_2$ and $A_3$ ($B_1$, $B_2$ and $B_3$) represents the meta-atom on the sublattice *A*(*B*).

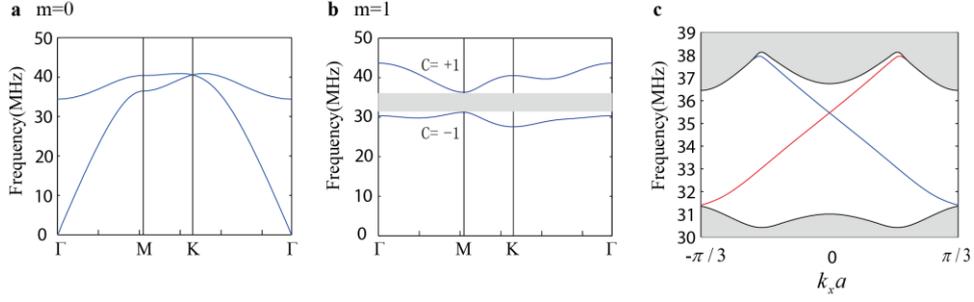

**Figure 2| Band structures and edge states. (a, b)** The bulk bands for *m*=0 **(a)** and *m*=1 **(b)**. The Dirac point exists at K for m=0 and the degeneracy is lifted for m=1 bands which possess nonzero Chern numbers as labeled in **(b)**. The band gap (gray region) between the two bands is nontrivial. The band structure for $m=-1$ is the same as that in **(b)** but the Chern numbers have opposite signs. **(c)** The projected band for *m*=1 along the *x* direction, where the gray area represents the projected bulk bands and the red and blue curves represent the edge states localized at the upper and lower boundaries of a ribbon of this system. The ribbon is periodic along *x* and truncated with an armchair boundary in the *y* direction. In the calculation, the intralayer and interlayer cable lengths are *a*=0.43m and *b*=2.06m, respectively. The wave speed is 0.66c and the cable loss is ignored.

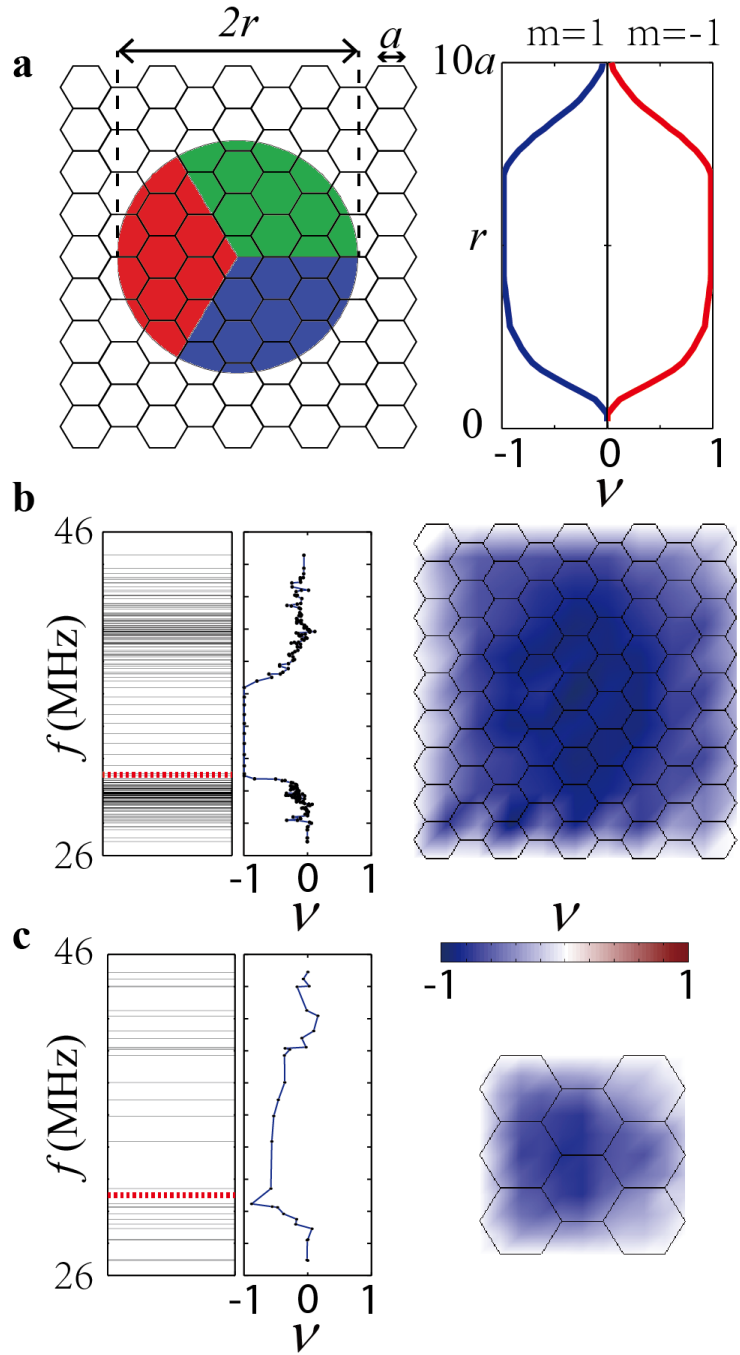

**Figure 3| Local Chern numbers.** (a) We use a circular computational domain for calculating the local Chern numbers at the sample center as a function of the computational domain radius for a $9\times9$ finite-size sample. The red, blue and green regions correspond to the *A, B,* and *C* sectors, respectively. (b) and (c) show the energy levels (left), local Chern numbers at the sample center (middle) and the local Chern number as a function of position (right) of $m=1$ states for $3\times3$ and $9\times9$ samples, respectively. In the calculations, the radii of their computational domain are set to

2.3*a* and 5*a* respectively. The black lines represent the eigenfrequencies. The dotted red line in **(b)** in the left panel marks *f*=31 MHz, which is set to the cutoff frequency in the local Chern number calculation. The middle panel shows how the local Chern number $\nu$ changes with the cutoff frequency, where the center of the computational domain is fixed at the center of the samples. Each point represents the value of $\nu$ by adding up all the eigenstates at and below this frequency.

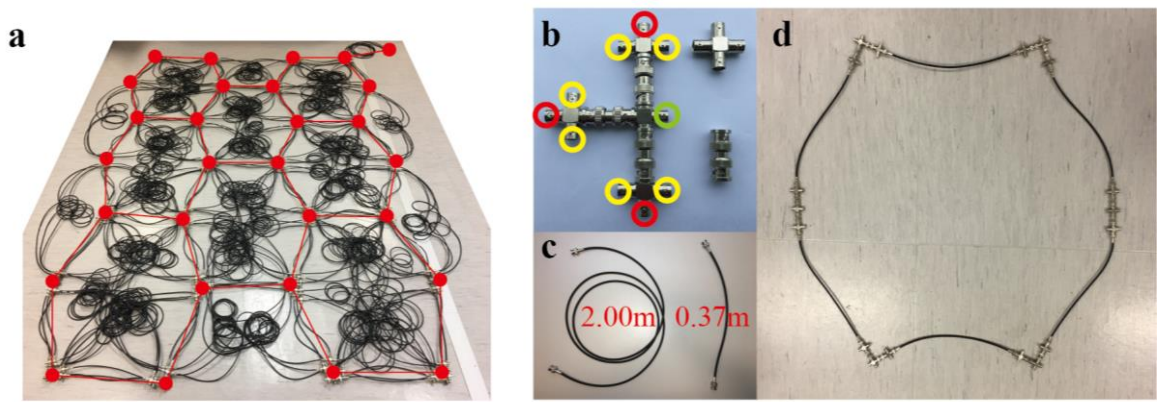

**Figure 4| The transmission line network.** **(a)** A photo of the experimental sample. The nodes are marked by red dots, which are connected by red lines to illustrate the 2D structure. **(b)** A node formed by connectors. The red, yellow and green circles mark the ports for connecting the intralayer, interlayer and measurement cables respectively. **(c)** 2.00 m and 0.37 m long cables. **(d)** A hexagon formed by cables and connectors.

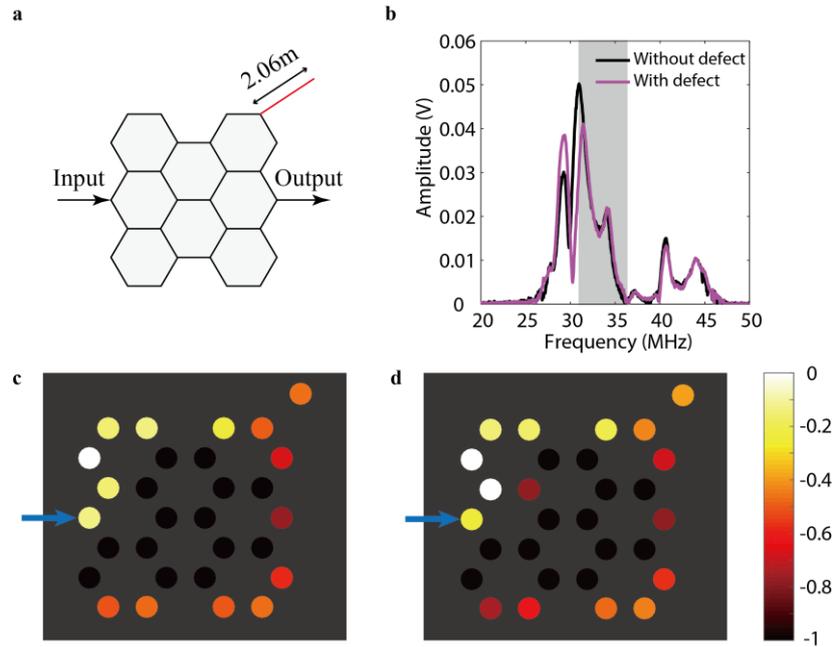

**Figure 5| Transmission spectra and edge state field distributions.** (**a**) The schematics of the experimental setup for measuring the transmission spectra of *m*=1 states (with a defect as represented by the red line). (**b**) The defect does not change the transmission spectra inside the nontrivial gap range (gray zone). (**c, d**) Experimentally measured (**c**) and numerically calculated (**d**) field patterns at 34.5 MHz in the nontrivial gap frequency range for *m*=1. The signal enters from the point marked by the blue arrow. The color of each node represents the value of $\ln|U/U_{max}|$, where $U$ is the voltage at each node, and $U_{max}$ is the maximum voltage among all nodes. The edge mode passes over corners and through the defect without backward scattering. The amplitude of the edge state decreases gradually due to the intrinsic loss in the cables.